\magnification=1200
\def\R{{\bf R}}
\def\next{\hfil\break\noindent}
\def\d{\partial}

\def\d{\partial}
\def\f#1#2{{\textstyle{#1\over #2}}}

\font\title=cmbx12

{\title 
\centerline{Cosmological spacetimes not covered }
\centerline{by a constant mean curvature slicing }}

\vskip 1cm

\noindent
{James Isenberg}
\next
{Department of Mathematics and }
\next
{Institute of Theoretical Science}
\next
{University of Oregon}
\next
{Eugene, OR 97403}
\next
{USA}

\vskip 10pt\noindent
{and}
\vskip 10pt\noindent
{Alan D. Rendall}
\next
{Max-Planck-Institut f\"ur Gravitationsphysik}
\next
{Schlaatzweg 1}
\next
{14473 Potsdam}
\next
{Germany
}

\vskip 1.5cm\noindent
{{\bf Abstract}} 

We show that there exist maximal globally hyperbolic solutions of the 
Einstein-dust equations which admit a constant mean curvature Cauchy surface, 
but are not covered by a constant mean curvature foliation. 

\vfil\eject

\noindent
{\bf 1. Introduction}

The most widely advocated candidate to be used as a choice of time in 
cosmological spacetimes, i.e., spacetimes which are 

\noindent
a) globally hyperbolic with a compact Cauchy surface, and

\noindent
b) solutions of Einstein's equations coupled to \lq reasonable\rq (see
Sect. 2) non-gravitational fields

\noindent
is that which is specified by a constant mean curvature (\lq CMC\rq )
foliation. Supporting this advocacy is the belief that a generic
cosmological spacetime admits a global CMC foliation which is unique
up to leaf-preserving diffeomorphism. While this belief may turn out
to be true, it was shown some years ago [1] that there are some
cosmological spacetimes which do not admit any CMC Cauchy surfaces.
Here we show further that there are also cosmological spacetimes
which admit CMC Cauchy surfaces but do {\it not} admit a global 
{\it foliation} by such hypersurfaces.

The spacetimes which we show have this property are quite special
\footnote*{Note that, thus far, the spacetimes which are known not to admit 
any CMC hypersurfaces are also special in that they are solutions of the
Einstein-dust field equations. No vacuum solutions are known to have this
property}: the matter fields in them are pressureless fluids, i.e. dust,
which are contrived to form \lq shell-crossing singularities\rq, and the 
spacetime fields are all invariant under a $T^2$-isometry group. While the
latter feature is merely a convenience, it may well be that the formation
of a shell-crossing singularity is crucial to the failure of the CMC
slicing to cover the spacetime. This issue needs further study. In any case,
we believe it is useful to know that there are cosmological spacetimes in
which CMC-based time exists for a while, but does not cover the whole
globally hyperbolic spacetime.

Our results here, to an extent, build on previous work by one of us[12].
In that work, it was shown that there are cosmological spacetimes (with
$T^2$ isometry and satisfying the Einstein-dust equations) which, starting
from initial data with finite constant mean curvature
$\tau_0$, develop a (shell-crossing) 
singularity arbitrarily soon - $\tau_0+\epsilon$ - in CMC time, provided
the solution exists that long. What we show here is that, while the 
singularity stops the CMC foliation from continuing past $\tau_0+\epsilon$, 
for some (generic?) choices of initial data, the spacetime continues to 
develop globally hyperbolically in regions which are spatially distant from 
the singularity. Hence, there are spacetime regions which are left uncovered 
by the CMC slicing. Note that while examples of spacetimes with 
incomplete CMC foliations can be constructed trivially by cutting pieces out
of certain spacetimes (see, for example, p. 472 of [13]), such examples are
not maximal globally hyperbolic developments of initial data, as are our
examples described here.
   
To prove the existence of spacetimes with these properties, we first
(Sect. 2) briefly discuss cosmological spacetimes satisfying the 
Einstein-dust equations, noting especially results concerned with 
stability, and we describe the special form of the fields and some of the 
field equations if we assume that the spacetimes are $T^2$ symmetric.
Next, in Sect. 3, we describe the special sets of initial data which were
used in [12] to  produce singularities in arbitrarily 
short CMC time. Then in Sect. 4, we specify the \lq quiet zone\rq\ 
restrictions on the data which we use to produce our present results, and we 
state and prove the results in Theorem 3 in that section.

\vskip .5cm\noindent
{\bf 2. Cosmological spacetimes with dust}

As noted earlier, a spacetime $(M^4,g_{\alpha\beta},\Psi)$ with metric 
$g_{\alpha\beta}$ and non-gravitational fields $\Psi$ is called
{\it cosmological} if $(M^4,g)$ is globally hyperbolic with compact Cauchy 
surface, and if $g_{\alpha\beta}$ and $\Psi$ together satisfy the coupled
Einstein-matter system:
$$\eqalignno{
G_{\alpha\beta}[g]&=8\pi T_{\alpha\beta}[g,\Psi]&(1a)                 \cr
{\cal F}[g,\Psi]&=0&(1b)}$$
where $T_{\alpha\beta}[g,\Psi]$ is the stress-energy tensor of $\Psi$
and $g_{\alpha\beta}$, and where ${\cal F}[g,\Psi]=0$ represents the field 
equations which the chosen field theory imposes on $g_{\alpha\beta}$
and $\Psi$. The globally hyperbolic condition implies that 
$M^4=\Sigma^3\times\R$ for some compact three-dimensional manifold 
$\Sigma^3$, and also implies that any spacelike embedding of $\Sigma^3$
in $(M^4,g)$ is a Cauchy surface[2]. Global hyperbolicity does not,
however, guarantee that the spacetime $(M^4,g_{\alpha\beta},\Psi)$ is
uniquely determined by initial data on a Cauchy surface. For this
property to hold, the PDE system (1) must be {\it hyperbolic} in an
appropriate sense[6]. This PDE hyperbolicity condition, along with an 
energy condition (weak, strong, or dominant)[7]) on the energy-momentum 
tensor, is what we mean by \lq reasonable\rq\ non-gravitational fields
in the definition of cosmological spacetimes.

Our focus in this paper is the Einstein-dust field theory, in which the
non-gravitational fields consist of a non-negative function $\mu$ (the
proper energy density of the matter) and a unit vector field $u^\alpha$
(the four-velocity of the matter). The stress-energy tensor for this
theory takes the form
$$T_{\alpha\beta}=\mu u_\alpha u_\beta,\eqno(2)$$
and the non-gravitational field equations are just those derived from 
the Bianchi identity
$$\eqalign{0&=\nabla_\alpha G^\alpha{}_\beta [g]           \cr
            &=\nabla_\alpha T^\alpha{}_\beta [g,\mu,u]}\eqno(3)$$
One readily verifies from (2) that the weak, strong and dominant energy
conditions are all satisfied by the Einstein-dust theory. In contrast,
it is {\it not} so easy to verify that the PDE system for this theory
is hyperbolic. The procedure of reduction to a symmetric hyperbolic system,
which straightforwardly shows hyperbolicity for the Einstein equations 
coupled to many matter fields, runs into difficulties in the case of dust.
(In contrast it does work for the Einstein-perfect fluid equations with
a reasonable equation of state and strictly positive pressure.) By
rewriting (1)-(3) in the equivalent form
$$\eqalignno{u^\gamma\nabla_\gamma G_{\alpha\beta}&=-\mu (\nabla_\gamma
u^\gamma)u_\alpha u_\beta&(4a)                                    \cr
u^\gamma\nabla_\gamma u_\beta&=0&(4b)                             \cr
\nabla_\gamma(\mu u^\gamma)&=0&(4c)}$$
Choquet-Bruhat is able to show that the Einstein-dust field equations are
hyperbolic in the Leray sense[3],[9].

It is important for our present purposes to know not only that one can 
determine the spacetime $(M^4,g_{\alpha\beta},\mu,u^\alpha)$ from knowledge of
initial data on a Cauchy surface $\Sigma^3_{\tau_0}$, but also that at least
for a finite (in time) spacetime neighbourhood $M^4_{(\tau_0-\lambda,
\tau_0+\lambda)}$ of $\Sigma^3_{\tau_0}$, the fields in this determined
spacetime 
$(M^4_{(\tau_0-\lambda,\tau_0+\lambda)}, g_{\alpha\beta}, \mu, u^\alpha)$
are {\it stable} under perturbations of the initial data. The theorem
which guarantees this stability for PDE systems which are hyperbolic in the 
sense of Leray was proven by Choquet-Bruhat in [4] using harmonic 
coordinates. Applied to the Einstein-dust equations (4) this result says the 
following.

\vskip 10pt
{\it Proposition 1. (Choquet-Bruhat [4]) Let $(M^4_0,g_{\alpha\beta}^0,
\mu_0,u^\alpha_0)$ be a cosmological spacetime satisfying the harmonically
reduced Einstein-dust equations and let
$(\hat M^4_0,\hat g_{\alpha\beta}^0,\hat\mu_0,\hat u^\alpha_0)$ be the 
restriction of this solution to a region $\hat M^4_0$ bounded between a pair 
of Cauchy surfaces in $M^4$. Let $(\hat g^0_{\alpha\beta}, \hat\mu_0, 
\hat u_0^\alpha)$ be contained in the Sobolev space $H_7(\hat M^4)$, and for 
a fixed Cauchy surface $\Sigma^3$ in $\hat M^4$, let the corresponding Cauchy
data $C_0$ be contained in $H_7(\Sigma^3)$. There exists an open set 
$W\subset H_7(\Sigma^3)$ with $C_0\in W$ such that every set of initial
data $C\in W$ generates a cosmological spacetime $(\hat M^4,
\hat g_{\alpha\beta}, \hat\mu, \hat u^\alpha)$ which satisfies the
harmonically reduced Einstein-dust equations, and also 
$(\hat g_{\alpha\beta}, \hat\mu, \hat u^\alpha)\in H_7(\hat M^4)$.
Moreover the mapping which takes a point of $W$ to the corresponding
solution is continuous (in fact differentiable) with respect to the 
$H_7$ topology.}

\vskip 10pt\noindent
The spacetimes that we show here have an incomplete CMC slicing are
Einstein-dust cosmological spacetimes, and in addition have $\Sigma^3=T^3$
and admit a $T^2$ isometry group which acts spatially. In a neighbourhood of 
some fixed CMC initial slice $\Sigma_0^3$, the metric for such a spacetime can
be written in the form[11]
$$g=-\alpha^2 dt^2+A^2 [(dx+\beta^1 dt)^2+a^2\tilde g_{AB}(dy^A+\beta^A dt)
(dy^B+\beta^B dt)]\eqno(5)$$
Here we use coordinates $(x,y^2,y^3,t)$, where $x$ and $y^A$ are spatial 
periodic coordinates on $T^3$, with $\d/\d y^A$ being Killing fields, and $t$ 
is a time coordinate (possibly, but not necessarily CMC). The functions 
$\alpha$, $A$, $\beta^1$ and $\beta^A$ are all functions of $x$ and $t$ 
(periodic in $x$), as is the two-dimensional unit determinant Riemannian
metric $\tilde g_{AB}$; the function $a$ depends on $t$ only. We also require 
that the density function $\mu$ depend only on $x$ and $t$, and that the 
matter velocity field $u^\alpha$ take the form $u=vA^{-1}\d/\d x+w\d/\d t$, 
with $v$ a function of $x$ and $t$ only and with $w$ determined by the 
condition $g_{\alpha\beta} u^\alpha u^\beta=-1$.

Substituting the metric (5) into the Einstein-matter
equations (1a), (2) and (3),
and using the assumptions just stated, one obtains an explicit formulation
of the Einstein-dust Cauchy problem for a class of $T^2$-symmetric 
cosmologies in terms of $\alpha$, $A$, $\beta^1$, $\beta^A$, $a$, $\mu$ and 
$v$. While we will need to use the local existence and stability results for
the full Cauchy problem - see Proposition 1 above - we shall only be working 
with a couple of the equations in explicit form. The two we will need - they
are constraints on the choice of initial data - are [12]:
$$\eqalignno{
\d^2/\d x^2(A^{1/2})&=-\f18 A^{5/2}[\f32 (K-\f13\tau)^2-\f23\tau^2+\sigma^2
+16\pi\mu (1+v^2)]&(6a)                              \cr
\d K/\d x&+3A^{-1}(\d A/\d x) K-\f13 A^{-3}\d/\d x (A^3\tau)-\lambda=8\pi A
\mu(1+v^2)^{1/2}v,&(6b)}$$
where $\tau$ is the mean curvature of the chosen initial data (generally
a function of $\alpha$, $\beta^1$, $\beta^A$, $a$ and their time derivatives),
where $K$ is an eigenvalue of the second fundamental form of the initial
data - specifically one has:
$$K=\f23\alpha^{-1}[a^{-1}\d a/\d t+\d\beta^1/\d x]+\f13\tau\eqno(7)$$
and where $\sigma$ and $\lambda$ are certain functions of the initial data 
which we need not specify, since they vanish for the cases we consider below.

\vskip .5cm\noindent
{\bf 3. Initial Data for Spacetimes with Finite CMC Slicing}

In our examples the mechanism which is envisaged as stopping the CMC slicing
from proceeding is the formation of \lq shell-crossing singularities\rq. It is
appropriate at this point to discuss this feature of dust spacetimes in
some more detail. The general idea is that self-gravitating dust particles
have a tendency to collide with each other, creating singularities. More
specifically, in a spacetime with an isometry group with two dimensional 
orbits, a shell of dust particles which are related to each other by the
symmetry moves in a coherent way. If two of these shells collide then the
intermediate shells are trapped between them, so that the matter density
is forced to blow up. Let us give a formal definition in the case of $T^2$
symmetry which is of interest here.

\vskip 10pt
{\it Definition. Consider a $T^2$-symmetric cosmological spacetime which is a 
solution of the Einstein-dust equations on which is defined a time coordinate 
$t$ with compact level surfaces whose range is $(t_-,t_+)$. Suppose that 
there is a time $t_0$ with $t_-<t_0<t_+$ and two distinct dust particles 
which move orthogonal to the group orbits with the property that the distance
between their world lines, as measured in the hypersurfaces $t$=const. tends
to zero as $t\to t_+$. Then we say that the spacetime develops a 
shell-crossing singularity as $t\to t_+$.}

\vskip 10pt\noindent
In the examples which we discuss in the following we will not prove that a
shell-crossing singularity develops. However the strategy of the proof is
guided by the idea that that is what is happening.  To construct these 
examples, we rely on the initial value formulation, so our aim is to find 
simple initial data sets for Einstein-dust cosmological spacetimes, which
are apt to lead to shell-crossing in the future, and which of course
satisfy the Einstein constraint equations. We do this using the conformal 
method [5], as adapted to the $T^2$ symmetric metric (5) and the 
corresponding matter variables discussed in Sect. 2. As in [5], we proceed as 
follows:

\vskip 10pt\noindent
-- Choose (for some fixed constant $t_0<0$)

\noindent
(i) $\tau=t_0$
\next
(ii) $\tilde g_{ab}=\delta_{ab}$ (so $\lambda=0$)
\next
(iii) $a=1$
\next
(iv) second fundamental form so that $\sigma=0$
\next
(v) $\tilde\mu (x)$ any smooth positive function on $S^1$ with
$$\tilde\mu (\pi-x)=\tilde\mu (x)\eqno(8a)$$
and
\next
(vi) $\tilde v(x)$ any smooth function on $S^1$ with
$$\tilde v(\pi-x)=-\tilde v(x)\eqno(8b)$$.

\vskip 10pt\noindent
-- Determine $\tilde K(x)$ by solving
$$\d\tilde K/\d x=8\pi\tilde\mu (1+\tilde v^2)\tilde v\eqno(9a)$$
and $A(x)$ by solving 
$$\d^2/\d x^2(A^{1/2})=-\f18 A^{5/2}[\f32 A^{-3}(\tilde K-\f13 t_0)^2
-\f23 t_0^2+16\pi A^{-4}\tilde\mu(1+\tilde v^2)]\eqno(9b)$$
-- Set
$$\eqalignno{
&\mu=A^{-4}\tilde\mu&(10a)                \cr
&v=\tilde v&(10b)                         \cr
&K=\f13 t_0+A^{-1}(\tilde K-\f13 t_0)&(10c)}$$
For a given choice of $t_0$, $\tilde\mu(x)$ and $\tilde v(x)$, this method 
produces a solution of the constraints so long as equations (9a) and (9b)
admit a solution $\tilde K$ and $A$. The reflection symmetry conditions
(8a)-(8b) guarantee that the right hand side of (9a) satisfies
$$\int_{S^1}\tilde\mu(1+\tilde v^2)\tilde v dx=0,$$
from which it follows that (9a) admits a solution (unique up to a 
constant). Sub and supersolution techniques [8] show that (9b) admits a 
solution as well, with that solution bounded below and above by the 
subsolution $A_-$ and the supersolution $A_+$ respectively:
$$\eqalignno{
A_-&=t_0^{-1}(24\pi B)^{1/2}&(11a)                    \cr
A_+&={\rm Max}\{t_0^{-1}(48\pi\|\tilde\mu\|_\infty(1
+\|\tilde v\|^2_\infty)^{1/2},
t_0^{-2/3}(\f92\|\tilde K-\f13 t_0\|_\infty^2)^{2/3}\}&(11b)}$$  
Here $B:={\rm Min}\ \tilde\mu>0$. 

Note that, a priori, $\beta^1$ and $\alpha$ may be chosen freely. If
one wishes to maintain a CMC foliation into the future and past, however, 
then $\alpha$ must be chosen to satisfy
$$\d^2\alpha/\d x^2+A^{-1}(\d A/\d x)(\d\alpha/\d x)=\alpha A^2
[\f32(K-\f13 t_0)^2
+\f13t_0^2+4\pi\mu(1+v^2)]-A^2\eqno(12)$$
Similarly, to maintain a fixed $x$ coordinate range, say $[0,2\pi]$,
one must restrict $\beta^1$. We shall presume that this has been done. 

So far, we have not encoded shell-crossing into the initial data. This
is readily done: one chooses a pair of points $x_1$ and $x_2$ in the interval
$(0,\pi)$, and one chooses the function $v$ so that $v(x_1)=1$ and $v(x_2)=-1$.
These data describe a pair of dust particles (among others) starting at $x_1$ 
and $x_2$ at $t_0$, with initial spatial velocities $+1$ and $-1$. If one uses
CMC time - $\tau (t)=t$ - then the equations of motion for these particles
are
$$\eqalignno{
dx/dt&=\alpha A^{-1} v (1+v^2)^{-1/2}-\beta^1&(13a)                \cr
dv/dt&=-A^{-1}\d\alpha/\d x (1+v^2)^{-1/2}+\alpha K v&(13b)}$$
{}From these equations, as well as from the Einstein-dust equations, one
obtains upper and lower bounds for changes in the velocities of the particles 
in finite CMC time [12]. Since these bounds are independent of $|x_2-x_1|$,
one expects that for sufficiently small $|x_2-x_1|$, shell crossings are 
inevitable.

Now, as a consequence of local existence theorems for the Einstein-dust 
PDE system, we know that for every choice of sufficiently smooth initial 
data, a cosmological spacetime consistent with those data - i.e., a
cosmological {\it development} of that data - exists for at least a
finite proper time into the future and into the past of the initial Cauchy 
surface at $t_0$. In the special case of $T^2$ -symmetric spacetimes, local
existence theorems in terms of CMC coordinates have been proved [11];
so, one knows in addition that there is, in every maximal development of
initial data constructed as above according to conditions (i)-(vi) and
equations (9)-(10), a finite CMC time slicing in a neighbourhood of the 
initial surface.

While one expects shell-crossing to stop a cosmological development from
proceeding in proper time, it is possible a priori that CMC time could proceed
to its limits - which are $0$ and $-\infty$ in a cosmological spacetime
with $\Sigma^3=T^3$ - in such a spacetime. As shown in [12], this is not
the case. Indeed one has

\vskip 10pt
{\it Proposition 2. [12] Let $t_0<0$ and $\epsilon>0$. There exist sets of
CMC ($\tau=t_0$) initial data for the Einstein-dust system such that in any 
development of the data, the CMC slicing which exists in a neighbourhood of
the initial surface cannot be extended past either $t_0+\epsilon$ or
$t_0-\epsilon$.}

\vskip 10pt\noindent
The basic idea of the proof is to use the Einstein-dust equations - including
the CMC equation (12) - to obtain controls on the geometric quantities
$\alpha$, $A$, and $K$ in a finite CMC time interval about $t_0$, and then 
argue that for $|x_2-x_1|$ small enough, the dust particles starting at
$x_1$ and $x_2$ (with initial velocities $+1$ and $-1$ as described above,
and with metric described by equation (13)) must intersect within CMC
time $t_0+\epsilon$, if the solution exists that long. An intersection of 
this type would violate regularity and so CMC time must stop. Similarly,
if one chooses $v(x)$ with $v(x_1)=-1$ and $v(x_2)=+1$, the intersection
occurs in the past, within $t_0-\epsilon$. For details, see [12].

We note two facts regarding this result which are important for our work in
the next section. First, we note that shell-crossing is essentially a 
local phenomenon. Specifically, it follows from the details of the proof
of Proposition 2 that regardless of what is happening outside of the interval
$[x_1,x_2]$, one can choose data inside $[x_1,x_2]$ which results in 
shell-crossing occurring arbitrarily soon. Second, it is a trivial
consequence of Proposition 2 as stated that given any sequence of positive
numbers $T_m$ converging to zero, one can choose a sequence of Einstein-dust
initial data $C_m$ such that in any spacetime development of $C_m$, CMC
slicing extends no further (in CMC time) than $t_0+T_m$.

\vskip .5cm\noindent
{\bf 4. Spacetimes Extending Past the CMC Slicing}

We now consider sets of initial data of the form described above, with 
shell-crossing built in, but with a \lq quiet region\rq\ $[x_-,x_+]$
far away from $[x_1,x_2]$. In this quiet region, we choose
$$\eqalignno{
\tilde\mu (x)&=1,\ \ \ x_-<x<x_+ &(14a)                  \cr
\tilde v (x)&=0,\ \ \ x_-<x<x_+ &(14b)}$$
It follows from (9a) that $(\d\tilde K/\d x) (x)=0$ for $x_-<x<x_+$, so
making a convenient choice of constant, we have $\tilde K(x)=\f13 t_0$
on that interval. Thus we find that, modulo the conformal factor $A^2$,
which solves (9b) on $S^1$, we have in $[x_-,x_+]$ initial data for a 
spatially flat Friedmann-Robertson-Walker cosmology.

Our claim is that for all sets of initial data of this type, regardless of
how soon shell-crossing occurs, $A$ is sufficiently well-behaved in
$[x_-,x_+]$ that in the maximal spacetime development of a given set of
such data, the domain of dependence of $[x_-,x_+]$ extends a finite proper
time into the future and into the past, and hence extends past the CMC 
slicing, if that is cut off sufficiently quickly. This is the content of our 
main result:

\vskip 10pt
{\it Theorem 3. There exist Einstein-dust cosmological spacetimes, maximal 
developments of Cauchy data, which contain a CMC Cauchy surface but cannot
be foliated by a CMC slicing.}

{\it Proof.} We fix an initial time $t_0\in (-\infty,0)$, and on the circle
we fix a pair of disjoint intervals\footnote*{Without loss of generality,
we assume that $[x_1,x_2]\cap [\pi-x_+,\pi-x_-]=\phi$ as well as
$[x_1,x_2]\cap [x_-,x_+]=\phi$} $[x_1,x_2]$ (the shell-crossing region)
and $[x_-,x_+]$ (the quiet region). We restrict attention to sets of data
which satisfy conditions (i)-(vi) in the above procedure for constructing
initial data and equations
(9)-(10), but with conditions (8a) and (8b) on $\tilde\mu (x)$
and $\tilde v(x)$ replaced by the conditions that:

\noindent  
a) $\tilde\mu (x)$ is a positive smooth function on $S^1$ with
$$\tilde\mu (x)=1,\ \ \ x_-<x<x_+\eqno(15a)$$
and
$$\tilde\mu (\pi-x)=\tilde\mu (x)\eqno(15b)$$

\noindent
b) $\tilde v(x)$ is any smooth function on $S^1$ with
$$\eqalignno{
|\tilde v(x)|&\le 1&(16a)                 \cr
\tilde v(x)&=0,\ \ \ x_-<x<x_+&(16b)      \cr
\tilde v(\pi-x)&=-\tilde v(x)&(16c)       \cr
\tilde v(x_1)&=1&(16d)}$$
and
$$\tilde v(\hat x)=-1\eqno(16e)$$
for some $\hat x\in (x_1,x_2)$.
As discussed above, we can solve (9a) for $\tilde K$, and we obtain that
$\tilde K=\f13 t_0$ on $[x_-,x_+]$.

Now in accord with the comments at the end of Sect. 3, it follows from the 
details of the proof of Proposition 2 in [12] that one can choose a sequence 
of initial data $C_m$ of the form just described - with $\tilde\mu (x)$ the
same for all $m$ - such that the CMC slicing in any development of $C_m$ 
extends no further than $t_0\pm T_m$. What we now wish to show is that $A$ 
and its derivatives restricted to $[x_-,x_+]$ are controlled uniformly, for 
all sets of data in this sequence $C_m$.

First, we recall the sub and super solutions (11) for $A$ on $S^1$. The sub 
solution is clearly the same for all sets $C_m$. For the super solution
$A_+$, while $\|\tilde\mu\|_\infty$ and  $\|\tilde v\|_\infty$ are 
independent of $m$, the quantity $\|\tilde K-\f13 t_0\|_\infty$ may not
be. However, it follows from (9a) that
$$\eqalign{
|\tilde K-\f13 t_0|&=8\pi|\int_{S^1}\tilde\mu(1+\tilde v^2)\tilde v|\cr
&\le 16\pi^2\|\tilde\mu\|_\infty(1+\|\tilde v\|_\infty^2)
\|\tilde v\|_\infty^2}\eqno(17)$$
which is independent of $m$. Hence we have upper and lower positive bounds 
for $A(x)$ for the entire sequence. Next, as detailed in [12], we use a
generalization of an estimate of Malec and \'O Murchadha [10] to argue
that $\d A/\d x$ is also uniformly bounded along the sequence.

We now note the form of the Lichnerowicz equation (9b), which expresses 
$\d^2 A/\d x^2$ in terms of $A$, $\tilde\mu$, $\tilde K$, $\tilde v$ and
$\tau$. We know from above that $A$ is uniformly bounded, and we have 
chosen $\tilde\mu$ and $\tau$ to be fixed (and bounded) independent 
of the sequence. The only functions that change as $m$ grows, and could
cause trouble for $\d^2 A/\d x^2$ are $\tilde v$ and $\tilde K$. However,
in $[x_-,x_+]$ we have $\tilde v=0$ and $\tilde K-\f13 t_0$ for all $m$;
it follows that {\it in} $[x_-,x_+]$ we have $\d^2 A/\d x^2$ uniformly
bounded along the sequence. Taking spatial derivatives of the
Lichnerowicz equation (9b), we may similarly argue that spatial derivatives
of $A$ of all orders are, at least in $[x_-,x_+]$, uniformly bounded.

With these controls on $A$, we have established uniform $C^\infty$ bounds
on the sequence $C_m$ of initial data, restricted to $[x_-,x_+]$. We may now 
apply the Cauchy stability statement of Proposition 1, from which we infer
the following: there exists a pair of finite positive numbers $\lambda$ and
$\mu$ (with $\lambda\le \f12 (x_++x_-)$) and there exists a fixed harmonic
coordinate system $(x, y^1, y^2, t)$ on the spacetime region
$$\hat M^4=(x_-+\lambda,x_+-\lambda)\times S^1\times S^1\times (t_0-\nu,
t_0+\nu)$$
such that for each set of initial data $C_m$, there is a spacetime
$(\hat M^4, g_m)$ with the properties

\noindent
(i) $(\hat M^4, g_m)$ is a solution of the Einstein-dust equations
\next
(ii) the metric $g_m$ induces the initial data $C_m$ on the hypersurface
$t=t_0$
\next
(iii) the metric coefficients of $g_m$ and $g_m^{-1}$ written in terms of
the coordinates $(x, y^1, y^2, t)$ are all $C^\infty$ uniformly bounded

In other words, because of the \lq quietness\rq\ built into each set of
initial data $C_m$ on the region $[x_-,x_+]$, the maximal spacetime 
development $g_m$ of each $C_m$ lasts at least for a common (harmonic)
time $\nu$. Note that we use harmonic coordinates here because the Cauchy
stability proposition is proven [4] using harmonic coordinates.

Let us now fix a point $(x_*,y^A_*)\in (x_-+\lambda,x_++\lambda)\times S^1
\times S^1$; for each metric $g_m$, we consider the timelike geodesic
$\gamma_m$ which is orthogonal to the initial surface
$$(x_-+\lambda,x_++\lambda)\times S^1\times S^1\times\{t_0\}$$
at the point $(x_*,y^A_*,t_0)$. Since $g_m$ and its inverse are both
well-behaved in harmonic coordinates, one readily determines that for each
$\gamma_m$, there is a positive number $L_m$ which measures the proper time
length of $\gamma_m$ from the point $(x_*,y^A_*,t_0)$ on the initial surface 
to the point where $\gamma_m$ leaves $\hat M^4$. Moreover, the uniform
boundedness of $g_m$ and $g_m^{-1}$ implies the existence of some $L>0$ such 
that $L_m>L$ for all $m$.

By construction, in any spacetime development $(M^4,g_m)$ of the initial data 
$C_m$, there is a CMC slicing in the neighbourhood of the initial 
surface, with the slicing lasting no longer than $T_m$ in CMC time and 
with $\lim_{m\to \infty} T_m=0$. If we consider the length ${\cal L}_m$ of 
$\gamma_m$ restricted to the CMC region, we obtain the formula
$${\cal L}_m=\int_{t_0}^{t_0+T_*}\alpha(\gamma(t))dt\eqno(18)$$
where $t$ is CMC time.

{}From the CMC equation (12) for $\alpha$ (which holds for all CMC 
$t\in [t_0,t_0+T_m]$) we note that at any point where $\alpha$ achieves
a maximum, we have
$$0\ge A^2\alpha [\f32(K-\f13 t)^2+\f13 t^2+4\pi\mu(1+v^2)]-A^2\eqno(19)$$
which implies (recalling that $t<0$) that
$$\alpha\le 3/t_0^2\eqno(20)$$
Hence from (18), we have 
$${\cal L}_m\le (3/t_0^2)T_m\eqno(21)$$
But $\lim_{m\to \infty} T_m=0$, so that $\lim_{m\to \infty} {\cal L}_m=0$. 
This tells 
us that for some sufficiently large value of $m$, we have in $\hat M^4$
(and in any development $\bar M^4$ containing $\hat M^4$) ${\cal L}_m<L_m$.
Thus in $\bar M^4$, the geodesic $\gamma_m$ leaves the region of CMC slices
before it passes out of the region with harmonic coordinates.

Note that the arguments of this section do not rely essentially on the
fact that the matter model is dust; a perfect fluid with pressure would
do just as well. The only point where the fact that the matter model is
dust is used is in applying Proposition 2. If an analogue of Proposition 
2 could be proved for a perfect fluid with pressure then a generalization
of Theorem 3 to that matter model would follow. It seems plausible that
an analogue of Proposition 2 does hold for a fluid with pressure, due to
the formation of shock waves, but the proof would be very different from
that in the case of dust. On the other hand, there are other forms of 
matter for which the analogue of Proposition 2 is false; this has been
proved for collisionless matter and wave maps in [11]

\vskip 10pt\noindent
{\bf Acknowledgements} One of us (JI) thanks the Max Planck Institute for
Gravitational Physics for its hospitality while this research was being
carried out. This research was supported by NSF grant 9308117 at Oregon.

\vskip .5cm\noindent
{\bf References}

\noindent
[1] Bartnik, R.: Remarks on cosmological spacetimes and constant mean 
curvature surfaces. Commun. Math. Phys. {\bf 117}, 615-624 (1988).
\next
[2] Budic, R., L. Lindblom, J. Isenberg, P. Yasskin: On the determination
of Cauchy surfaces from intrinsic properties. Commun. Math. Phys. {\bf 61}, 
87-95 (1978).
\next
[3] Choquet-Bruhat, Y.: Th\'eor\`emes d'existence en m\'ecanique
des fluides relativistes. Bull. Soc. Math. de France {\bf 86}, 155-175 (1958)
\next
[4] Choquet-Bruhat, Y.: The stability of the solutions of nonlinear 
hyperbolic equations on a manifold. Russ. Math. Surveys {\bf 29}, No. 2, 
327-335 (1974)
\next
[5] Choquet-Bruhat, Y., J. York: The Cauchy problem. In General Relativity
and Gravitation (ed. A. Held), Plenum (1980).
\next
[6] Courant, R, Hilbert, D.: Methods of mathematical physics II, chapter VI,
Wiley (1989).
\next
[7] Hawking, S, G. F. R. Ellis: The large-scale structure of spacetime,
Cambridge University Press (1973).
\next
[8] Isenberg, J.: Constant mean curvature solutions of the Einstein 
constraint equations. Class. Quantum Grav. {\bf 13}, 1819-1847 (1996).
\next
[9] Leray, J.: Hyperbolic differential equations, Princeton, Institute for
Advanced Study (1951).
\next
[10] Malec, E., N. \'O Murchadha: Optical scalars and singularity avoidance
in spherical spacetimes. Phys. Rev. D  {\bf 50}, R6033-R6036 (1994).
\next
[11] Rendall, A. D.: Existence of constant mean curvature foliations in
spacetimes with two-dimensional local symmetry. Commun. Math. Phys.
(to appear)
\next
[12] Rendall, A. D.: Existence and non-existence results for global constant 
mean curvature foliations. To appear in proceedings of the Second World 
Congress of Nonlinear Analysts (Athens, 1996)
\next
[13] Rendall, A. D.: Constant mean curvature foliations in cosmological
spacetimes. Helv. Phys. Acta {\bf 69}, 490-500 (1996)

\end